\documentclass[doublecol]{epl2}

\title{Skyrmion Clusters From Bloch Lines in Ferromagnetic Films}

\author{Dmitry A. Garanin$^1$, Eugene M. Chudnovsky$^1$, \and Xixiang Zhang$^2$}

\institute{$^1$
 Physics Department, Herbert H. Lehman College and Graduate School, The City University of New York,
 250 Bedford Park Boulevard West, Bronx, New York 10468-1589, USA\\
$^2$Physical Science and Engineering Division, King Abdullah University of Science and Technology (KAUST),
Thuwal 23995-6900, Saudi Arabia}

\abstract{
Conditions under which various skyrmion objects emerge in experiments on thin magnetic films remain largely unexplained. We investigate numerically centrosymmetric spin lattices in films of finite thickness with ferromagnetic exchange, magnetic anisotropy, and dipole-dipole interaction. Evolution of labyrinth domains into compact topological structures on application of the magnetic field is found to be governed by the configuration of Bloch lines inside domain walls. Depending on the combination of Bloch lines, the magnetic domains evolve into individual skyrmions, biskyrmions, or more complex topological objects. While the geometry of such objects is sensitive to the parameters, their topological charge is uniquely determined by the topological charge of Bloch lines inside the magnetic domain from which the object emerges.}

\begin{document}

\maketitle

\section{Introduction}
Skyrmions were initially proposed in nuclear physics as models of elementary particles \cite{SkyrmePRC58,Polyakov-book} and later entered condensed matter physics in application to topological defects in ferro- and antiferromagnetic films \cite{BelPolJETP75,Lectures,Brown-book}, Bose-Einstein condensates \cite{AlkStoNat01}, quantum Hall effect \cite{SonKarKivPRB93,StonePRB93}, anomalous Hall effect \cite{YeKimPRL99}, and liquid crystals \cite{WriMerRMR89}. Recent years have seen an explosion of theoretical and experimental research on magnetic skyrmions that has been largely fueled by the prospect of developing topologically protected data storage.

In condensed matter theory, skyrmions emerge as topologically protected solutions of the continuous-field Heisenberg exchange model in two dimensions \cite{Lectures}, with the energy ${\cal{H}} = \frac{1}{2}\int d^2 r (\partial {\bf s}/\partial r_\alpha)^2$, where ${\bf s}$ is a three-component spin field of unit length and $r_\alpha = x,y$. An arbitrary spin field configuration, ${\bf s}({\bf r})$, is characterized by the topological charge,
\begin{equation}
Q = \frac{1}{4 \pi}\int dx dy \: {\bf s}\cdot \frac{\partial {\bf s}}{\partial x} \times\frac{\partial {\bf s}}{\partial y}
\label{Q}
\end{equation}
that takes values $Q = 0, \pm 1, \pm 2$, and so on.  They describe topologically different non-singular mappings of the $(s_x,s_y,s_z)$ unit sphere onto the $(x,y)$ geometrical space.

\begin{figure}
\includegraphics[width=88mm]{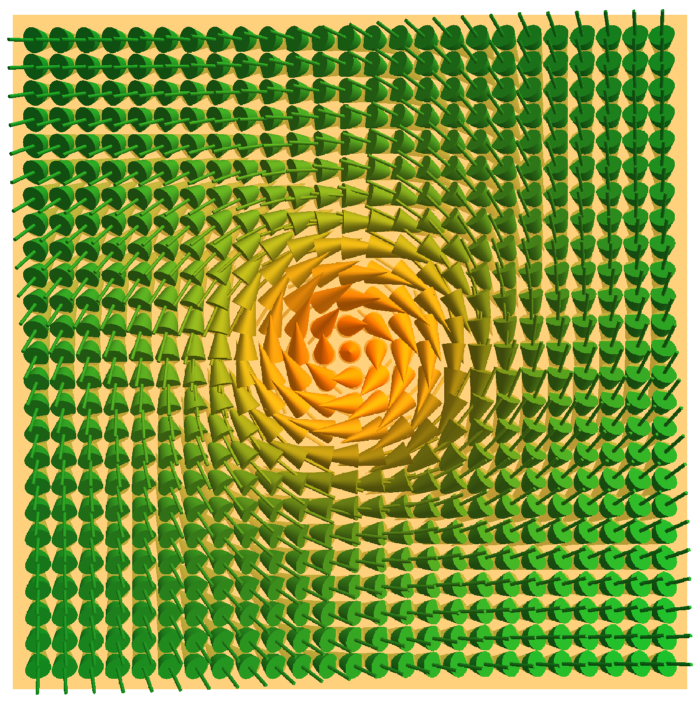}
\caption{Spin configuration of a Bloch skyrmion with $Q = 1$ in the purely exchange model.}
\label{Skyrmion}
\end{figure}

\begin{figure}
\includegraphics[width=88mm]{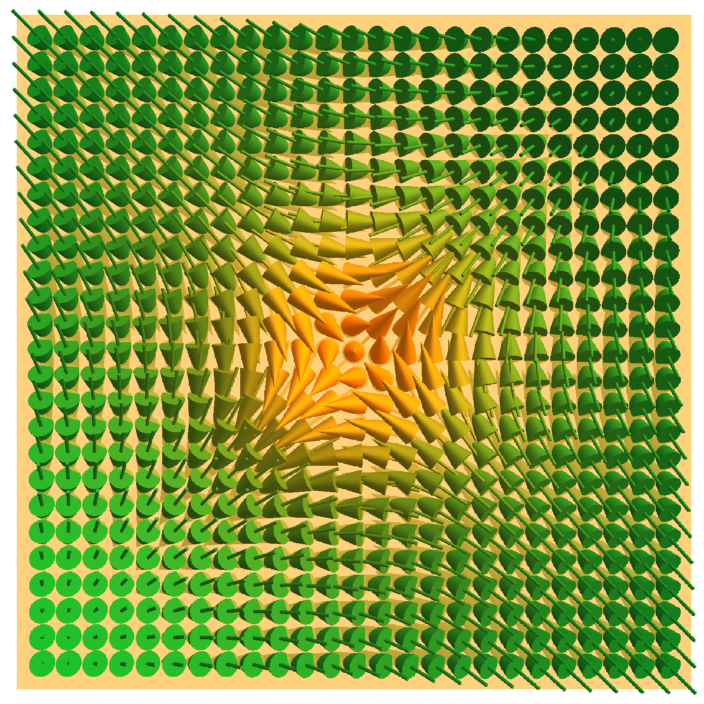}
\caption{Spin configuration of an antiskyrmion with $Q = -1$ in the purely exchange model.}
\label{Antiskyrmion}
\end{figure}

Examples of skyrmion and antiskyrmion are shown in Figs.\ \ref{Skyrmion} and \ref{Antiskyrmion}. If the spins are directed up at the center of the (anti)skyrmion and down at infinity, then $Q = 1$ corresponds to a skyrmion, while $Q = -1$ corresponds to an antiskyrmion. In this case that we will consider throughout the Letter, the topological charge can be figured out by going counter-clockwise around the (anti)skyrmion and looking at the rotation of the transverse magnetization. This rotation divided by $2\pi$ yields the topological invariant $Q$. The energy of the (anti)skyrmion does not depend on its chirality $\phi$ that for the skyrmion is the constant angle between the transverse magnetization and the radial direction. The Bloch (chiral) skyrmion shown in Fig.\ \ref{Skyrmion} has $\phi=\pm \pi/2$, whereas the N\'eel (achiral) skyrmion has $\phi=0$ or $\phi=\pi$. In general, if all spins in the (anti)skyrmion are rotated by a constant angle about the $z$ axis, its energy does not change. The energies of the skyrmion and antiskyrmion within the simple exchange model above are the same. Greater $|Q|$ describe multiskyrmion configurations. Dynamical conservation of the topological charge $Q$ is due to the scale invariance of the continuous-field exchange model that results in the (anti)skyrmion energy independent of its size. Violation of the scaling by the discreteness of the crystal lattice leads to the skyrmion collapse \cite{CCG}.

In this letter we study topological objects with different values of $Q$ generated by the magnetic field in the transverse ($z$) direction in the presence of the uniaxial transverse magnetic anisotropy and dipole-dipole interaction (DDI). This is the minimal model, studied throughout the Letter, in which skyrmions can be stabilized in spite of the lattice effect. In addition, Dzyaloshinskii-Moriya interaction (DMI), pinning, etc. can also stabilize lattice skyrmions in accordance with observations. Antiskyrmions in a 2D exchange model with DMI have been recently studied in Ref.\ \cite{Koshibae2016}.
\begin{figure}
\centerline{\includegraphics[width=88mm]{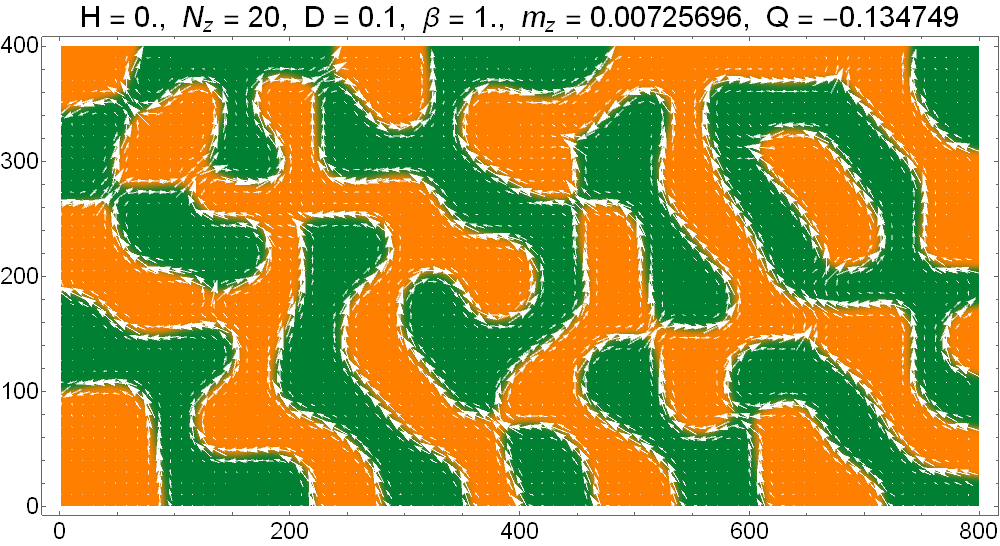}}
\centerline{\includegraphics[width=88mm]{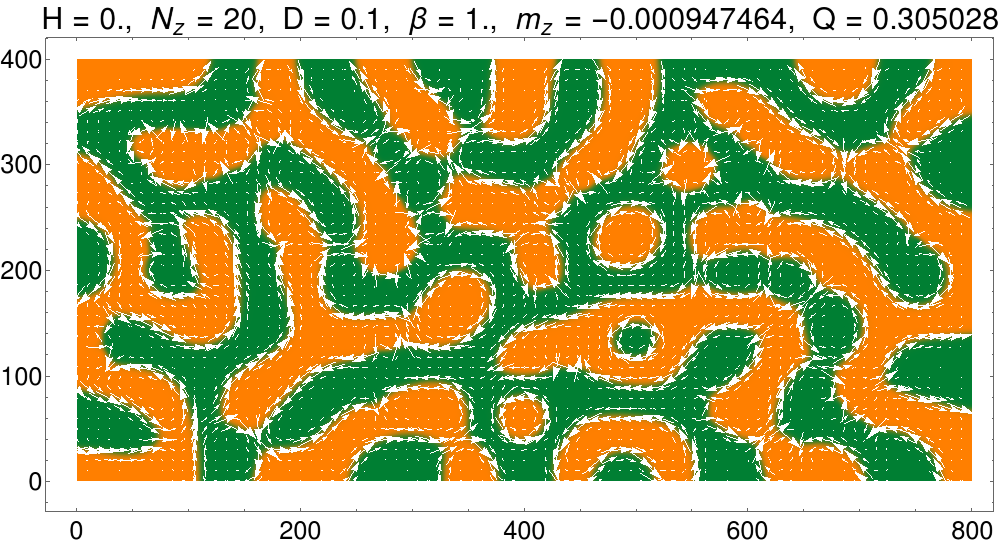}}
\caption{Typical pattern of laminar domains obtained numerically by relaxation from the initial state in a lattice model of a 800 $\times$ 400 film of 20 atomic layers. Domains with opposite directions of magnetization perpendicular to the film are color coded (orange/green being up/down). White arrows show in-plane spin components. Magnetization at the center of the domain wall separating two domains is parallel to the film. The reversal of the white arrows occuring as one moves along the domain wall indicates the presence of a Bloch line running through the domain wall perpendicular to the film. (Upper panel) Domain structure obtained from the initial state collinear along the $y$ axis. (Lower panel) The initial state with randomly oriented spins results in many more Bloch lines.}
\label{laminar}
\end{figure}
\begin{figure}
\includegraphics[width=88mm]{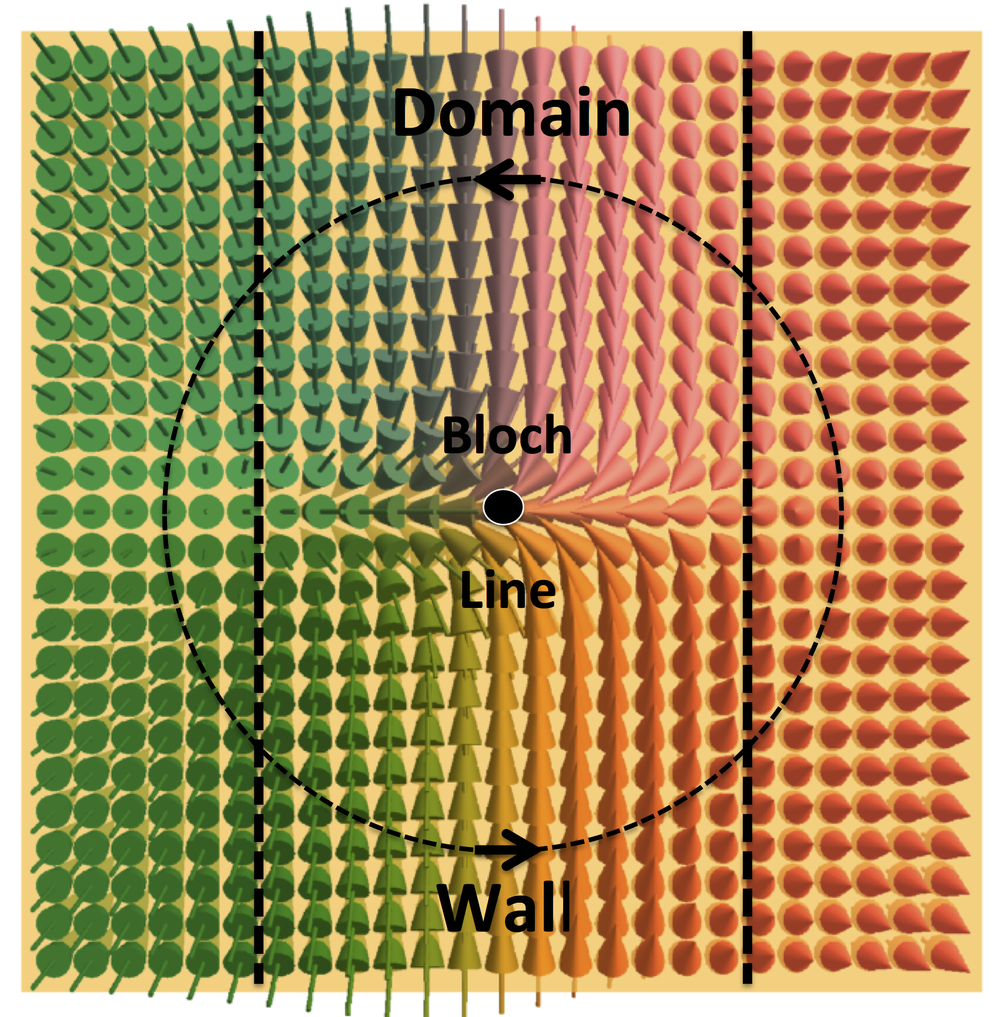}
\caption{Cross-section of a ferromagnet with the Bloch line inside the domain wall separating two regions (green and yellow) of opposite magnetization.  The direction of the spin reverses from down (green) to up (yellow) as one moves form the left to the right across the Bloch wall. The Bloch line shown by a solid circle at the center runs through the domain wall that is perpendicular to the picture. The rotation of the spin from one domain to the other occurs in different directions above and below the Bloch line.}
\label{BL}
\end{figure}

In a typical experiment, the initial ($B=0$) magnetic state of a ferromagnetic film with perpendicular magnetic anisotropy consists of a laminar structure of stripe domains shown in Fig.\ \ref{laminar}. Upon application of the transverse field, the stripe domains break into compact magnetic structures that may or may not be topologically protected. They have been studied experimentally and theoretically in numerous recent papers, see, e.g., reviews: Refs. \cite{Nagaosa2013}  and \cite{Klaui2016}. Besides individual skyrmions, stable biskyrmions \cite{Yu-2014,Zhang2016} and multiskyrmion clusters \cite{PNAC2016,Muller2017,Loudon2017} have been observed.

In non-centrosymmetric crystals with DMI the chiral skyrmions repel each other \cite{Leonov-NJP2016}, thus, inhibiting aggregation of skyrmions. It has been recently demonstrated, however, that the atraction between chiral skyrmions can occur due to a non-uniform background magnetization \cite{Leonov-JPCM2016,Loudon2017}. Such attraction also can result from a strong next-nearest-neighbor exchange interaction in the absence of DMI \cite{Leonov-NatCom2015,NNN-2017}). It should be noticed, however, the generic 2D exchange model possesses stable topological solutions with $Q > 1$ that are not clusters of skyrmions with $Q = \pm 1$.

In this paper we demonstrate that in centrosymmetric crystals skyrmions and skyrmion-like objects with $Q > 1$ naturally evolve from another type of topologically protected objects -- Bloch lines inside domain walls \cite{Lectures}, see Fig.\ \ref{BL}. Topological protection of the Bloch line comes from the rotation of the spin by $\pm 2\pi$ when going along a closed path of diameter large compared to the domain wall width (shown by the dash line in Fig.\ \ref{BL}) around the Bloch line. On crossing the Bloch line when moving in the plane of the domain wall the spin rotates by either $\pi$ or $-\pi$. This allows one to assign to the Bloch line a conserved topological number, $N_B = \pm 1$.

In the following, we numerically study the evolution of the stripe domain structure leading to the creation of skyrmion objects by increasing the applied field by small steps in the negative $z$ direction. In addition, we compute spin configurations obtained by relaxation from the initial condition in form of magnetic bubbles with a preset winding number (the topological invariant $Q$).

\section{Methods}
We study three-component spins in a ferromagnetic film of finite thickness with a cubic lattice and the energy given by
\begin{eqnarray}
{\cal{H}} & = & -\frac{J}{2}\sum_{ij} {\bf s}_i\cdot {\bf s}_j - B\sum_i s_{iz} - \frac{D}{2}\sum_i s_{iz}^2 \nonumber \\
& - & \frac{E_D}{2}\sum_{ij}\phi_{ij,\alpha\beta}s_{i\alpha}s_{j\beta}
\label{Hamiltonian}
\end{eqnarray}
with
\begin{equation}
\phi_{ij,\alpha\beta} \equiv a^3r_{ij}^{-5}\left(3r_{ij,\alpha}r_{ij,\beta} - \delta_{\alpha\beta}r_{ij}^2\right).
\end{equation}
Here $i,j$ denote lattice sites, with $r_{ij}$ being the distance between the $i$-th and the $j$-th site, $a$ is the lattice parameter, $\alpha,\beta = x,y,z$ denote spin components, $J$ is the exchange constant, $D$ is the constant of perpendicular magnetic anisotropy, and $E_D = \mu_0 M_0^2a^3/(4\pi)$ is the strength of the DDI (with $\mu_0$ being the magnetic permeability of vacuum and $M_0$ being the magnetization). The strength of the DDI relative to the anisotropy is given by the parameter $\beta \equiv D/(4\pi E_D)$.

In the presence of interactions that are weak compared to the exchange, the numerical work becomes inhibited by the required large system size and slow convergence. To speed up the computation, one can rescale the problem to another lattice constant $b>a$ by first rewriting the energy in the continuous approximation and then discretising it again. The rescaled model with the parameters
\begin{equation}
J'=\frac{b}{a}J,\quad B'=\frac{b^{3}}{a^{3}}B,\quad D'=\frac{b^{3}}{a^{3}}D,\quad E'_{D}=\frac{b^{3}}{a^{3}}E_{D}
\end{equation}
has two advantages: 1) smaller number of mesh points $N_{\alpha}'=(a/b)N_{\alpha}$; 2) smaller mismatch between $J$ and the other parameters that results in a faster  convergence.
\begin{figure*}
\centerline{\includegraphics[width=130mm]{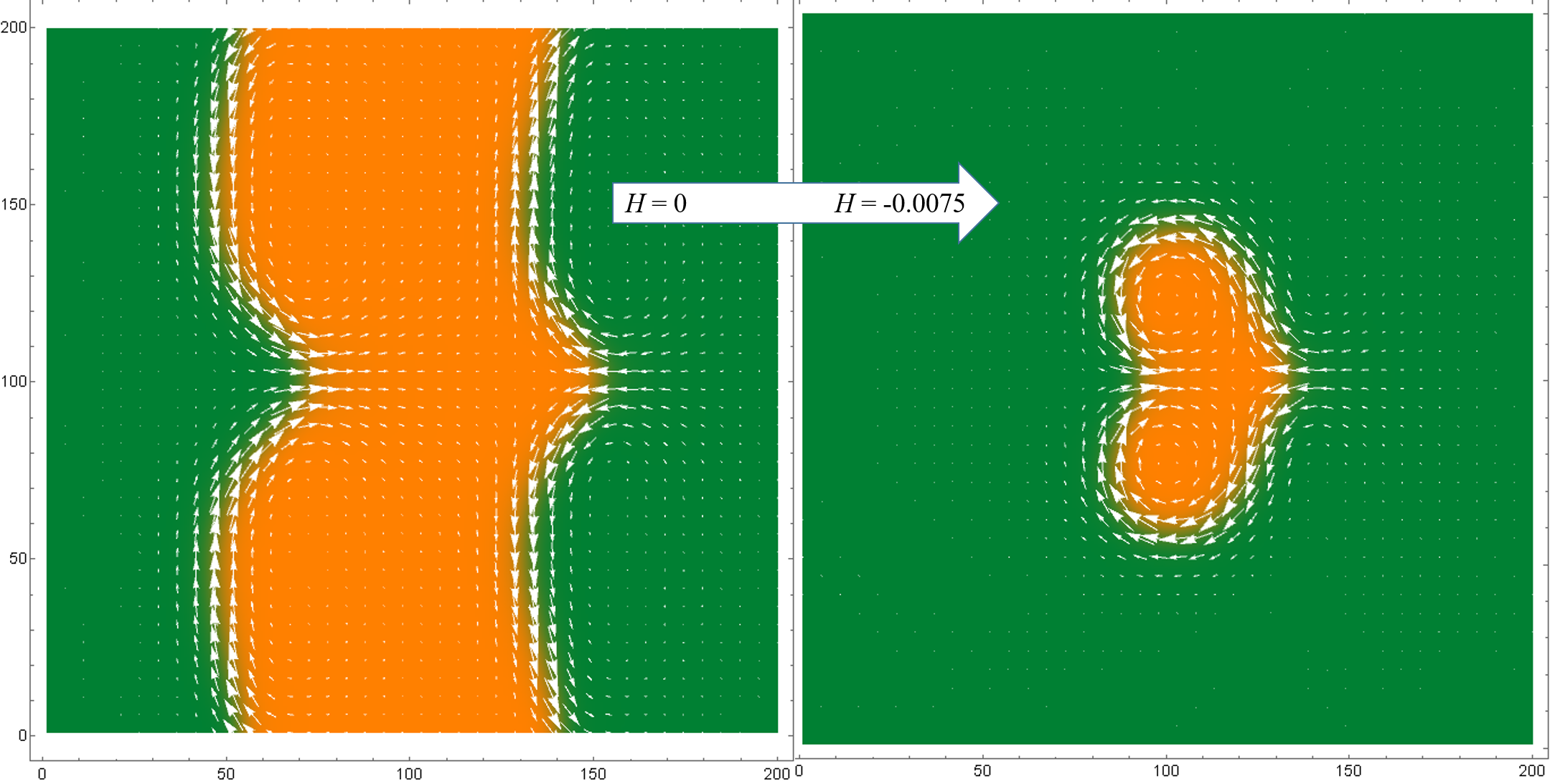}\includegraphics[width=16.5mm]{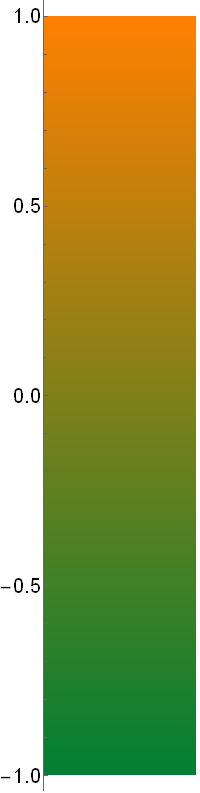}}
\caption{An example of the evolution of stripe domains with Bloch lines on application of the field. Perpendicular spin component $m_z$ is shown in orange/green corresponding to positive/negative $m_z$. White arrows show the in-plane spin components. On increasing the field applied normally to the $200\times 200\times 10$ film, the domain  transforms into a heart-shape bubble configuration with $Q = 1$. This heart bubble can transform into a common skyrmion with $Q = 1$ by annihilation of BLs with different rotation. }
\label{Combined-1}
\end{figure*}
\begin{figure}
\centerline{\includegraphics[width=80mm]{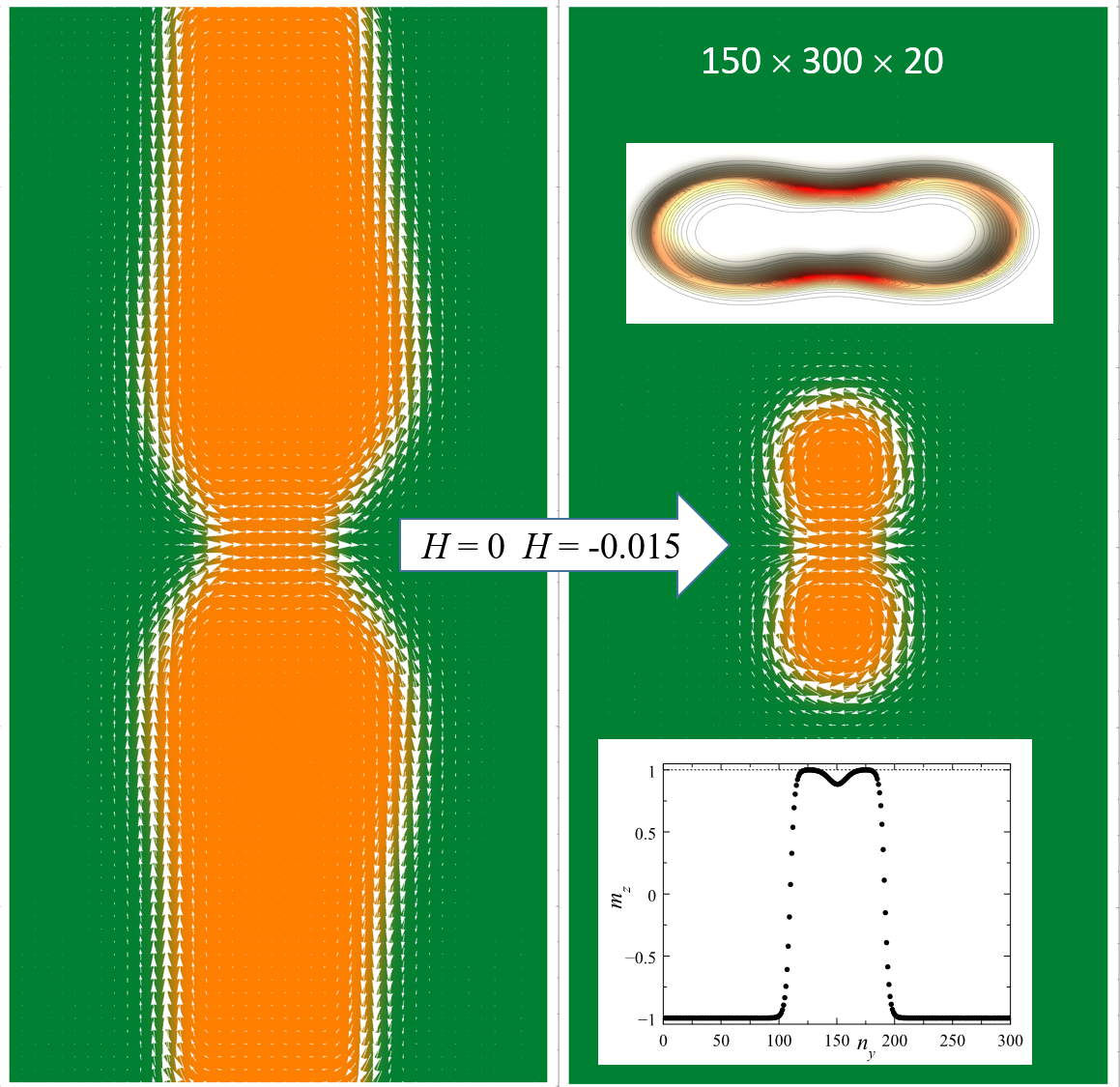}}
\centerline{\includegraphics[width=80mm]{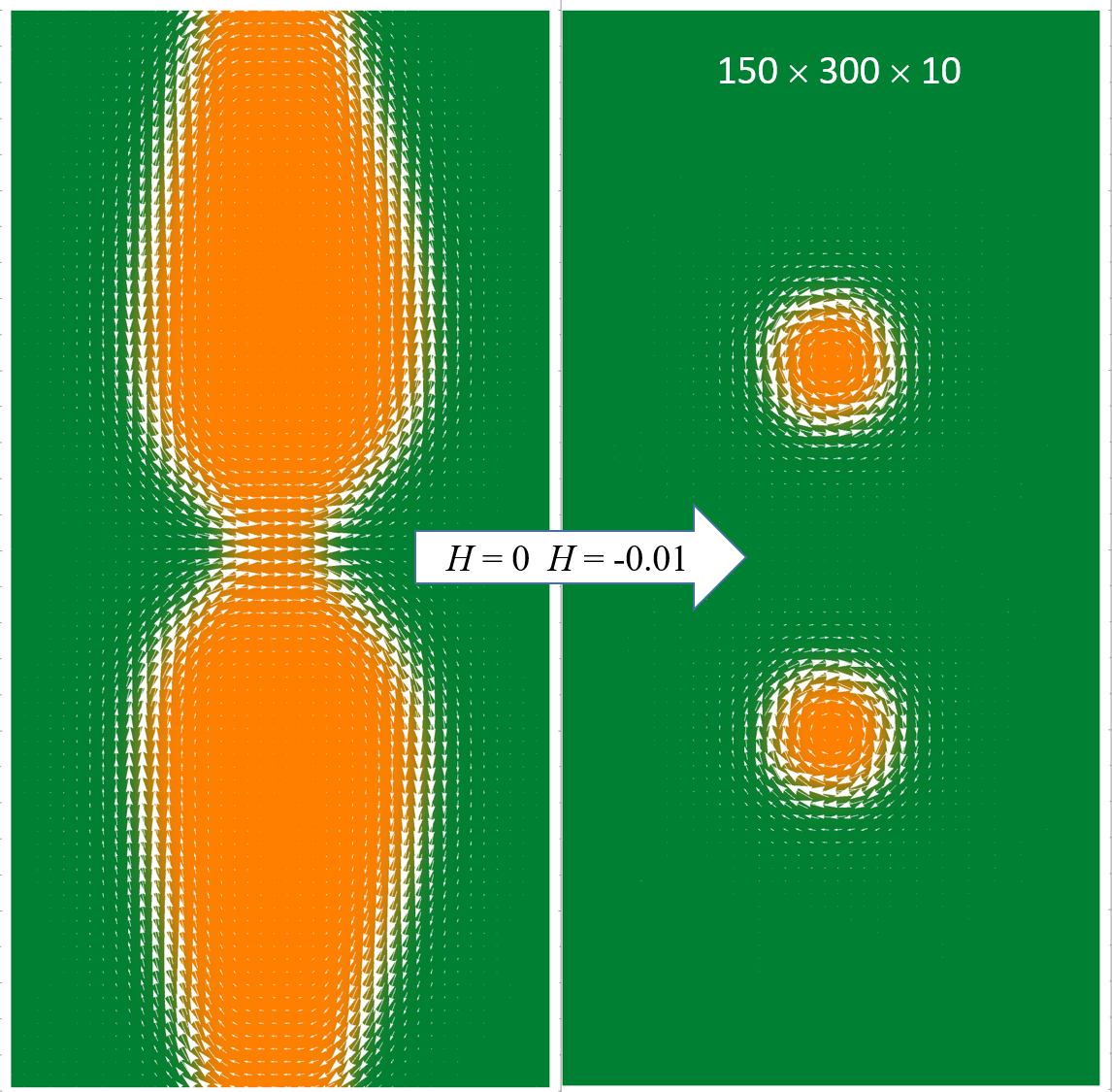}}
\caption{Evolution of stripe domains with Bloch lines on application of the field. (Upper panel) A longer domain in a $150\times 300\times 20$ film evolves into a tightly-bound biskyrmion with $Q = 2$. (Inset) Density of the topological charge $q_S$ [the integrand of Eq.\ (\ref{Q})] for the biskyrmion, zoomed in and rotated by 90 degrees. (Inset) The magnetization profile in the biskyrmion. (Lower panel) In a thinner $150\times 300\times 10$ film the biskyrmion is unstable and splits into two skyrmions because of a stronger effective DDI between spin columns at short range that favors repulsion between the skyrmions.
}
\label{Combined-3}
\end{figure}
 The maximal size used in our numerical work was $N_x \times N_y = 400 \times 800$ and $N_z$ between 10 and 100. The latter range of the thickness of the film in terms of the number of atomic layers is typical in experiments on skyrmions. In most cases the ratio $D/J =0.1$ (with $J = 1$ in all computations) was chosen to provide easily readable illustrations of the emerging physical objects. In real systems this ratio is smaller but the results remain qualitatively the same and can be obtained by rescaling, as shown above.  In most cases (except where specified differently) $\beta = 1$ was chosen to reflect situation that occurs in experiments.

We compute minimum-energy configurations of spins in large atomic lattices corresponding to films of finite thickness. Our numerical method combines sequential rotations of spins ${\bf s}_i$ towards the direction of the local effective field, ${\bf B}_{{\rm eff},i} = -\partial {\cal{H}}/\partial{\bf s}_i$, with the probability $\alpha$, and the energy-conserving spin flips (so-called {\it overrelaxation}), ${\bf s}_i \to 2({\bf s}_i\cdot {\bf B}_{{\rm eff},i}){\bf B}_{{\rm eff},i}/ B_{{\rm eff},i}^2-{\bf s}_i$, with the probability $1-\alpha$. The parameter $\alpha$ plays the role of the effective relaxation constant of the method. We use the value $\alpha=0.03$ that provides a faster convergence and deeper energy minimization than $\alpha=1$ for frustrated systems\cite{garchupro13prb}. The dipolar part of the effective field takes the longest time to compute. It has been done by the fast Fourier transform in the whole sample as one program step. Since the dipolar field is much weaker than the exchange, several cycles of spin alignment can be done before the dipolar field is updated, which increases the computation speed. For thin films that are studied here, the magnetization inside the film is nearly constant along the direction perpendicular to the film. Thus one can make the problem effectively two-dimensional by introducing the effective DDI between the columns of parallel spins. This greatly speeds up the computation. In all cases the total topological charge of the cluster, $Q$, has been computed numerically using the lattice-discretized version of Eq.\ (\ref{Q}). We used Wolfram Mathematica with compilation. The main
operating computer was a 20-core Dell Precision T7610 Workstation.

In the resulting plots we represent the transverse component of lattice spins $s_z$ by the color mapping: $s_z=1$ orange and $s_z=-1$ green. The in-plane components of the spins $s_x$ and $s_y$ are shown by white arrows. In addition, we show the density of the topological charge $q_S$ [the integrand of Eq.\ (\ref{Q})] with the color coding red/blue corresponding to positive/negative values.

\section{Results}
In what follows we consider evolution of a single stripe domain in a confined geometry on application of the field, $H = B/J$. Our main discovery is that the outcome, besides being sensitive to the parameters, depends drastically on the configuration of Bloch lines inside the domain walls. One of the scenarios is shown in Fig.\ \ref{Combined-1}. It describes evolution of a magnetic domain with Bloch lines into a heart-shaped cluster with $Q = 1$. Note that even in a zero field, BLs in the presence of the DDI make the domain wall non-straight.

Fig.\ \ref{Combined-3} shows the evolution of a stripe-domain structure with another type of rotation in Bloch walls that results into a tightly-bound biskyrmion, $Q = 2$, or two skyrmions (each having $Q=1$), depending on the film thickness that controls the effective DDI between the spin columns. While the exchange and the anisotropy favor aggregation of different skyrmions into one and decreasing their size to lower the energy, the DDI favors repulsion of skyrmions and increasing their size. In thinner films, the effective DDI is stronger at the short range that leads to instability of biskyrmions. The biskyrmion with $Q=2$ in  Fig.\ \ref{Combined-3}-left is identical to biskyrmions recently observed in experiments \cite{Yu-2014,Zhang2016}. Inset in Fig.\ \ref{Combined-3} shows the density of the topological charge $q_S$  in the biskyrmion. It is everywhere positive (red color) and localized in the biskyrmion  wall.

\begin{figure}
\centerline{\includegraphics[width=80mm]{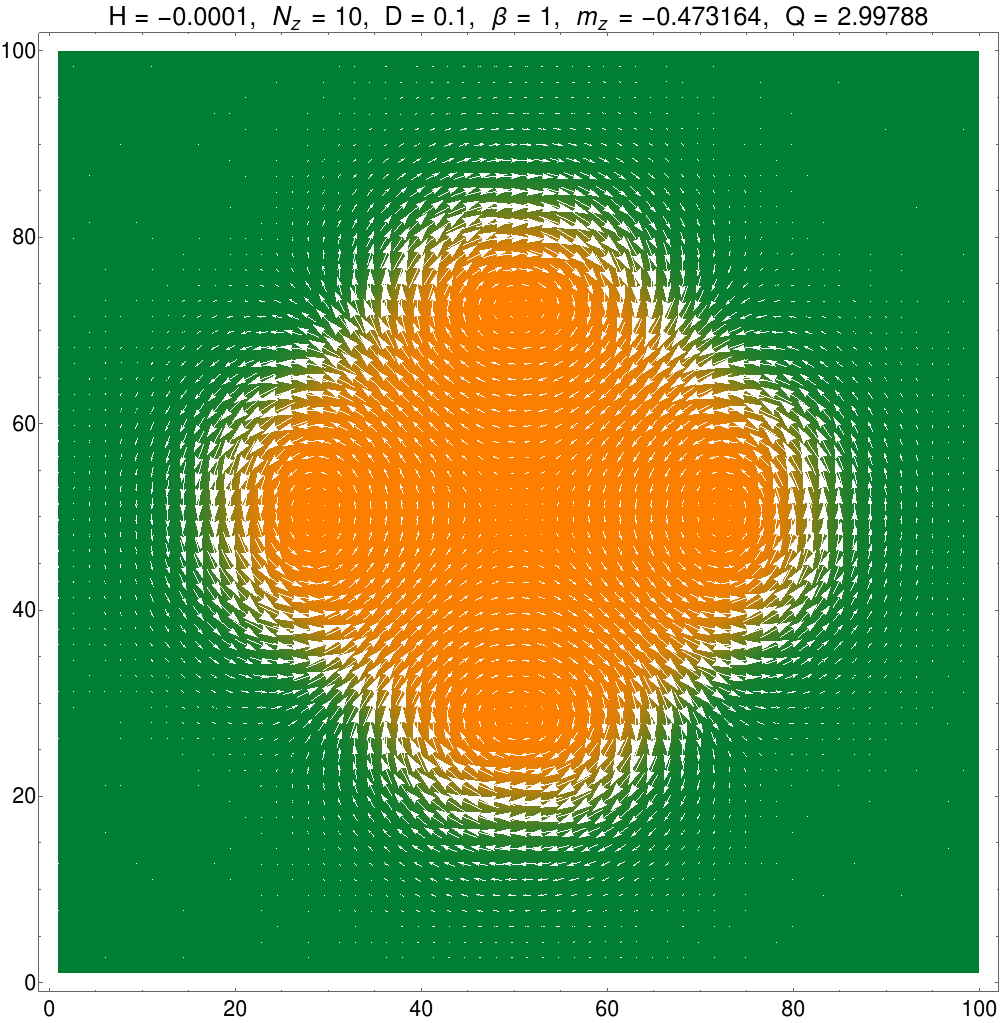}}
\centerline{\includegraphics[width=70mm]{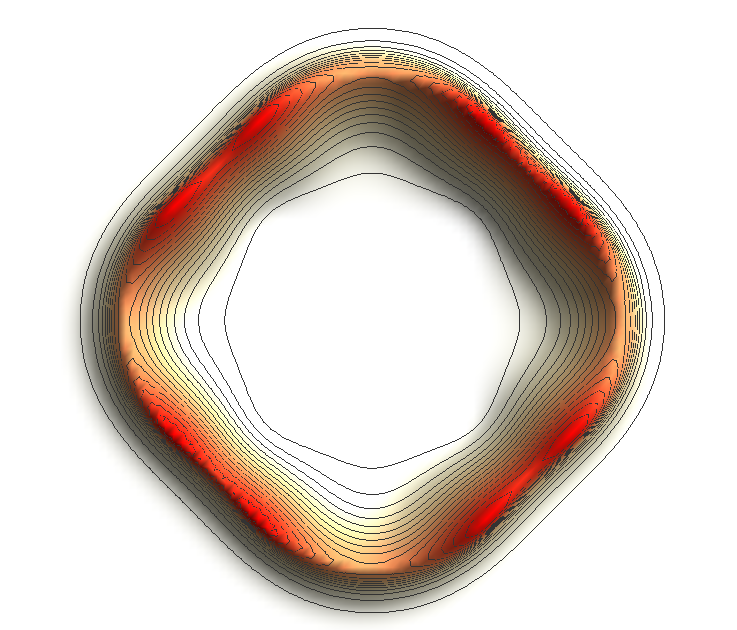}}
\vspace{0.5cm}
\caption{Skyrmion rosette with $Q = 3$ in a $100\times 100\times 10$ film. (Upper panel) $m_z$; (Lower panel) density of the topological charge $q_S$. Red/blue codes positive/negative $q_S$.}
\label{rosette}
\end{figure}

\begin{figure}
\centerline{\includegraphics[width=80mm]{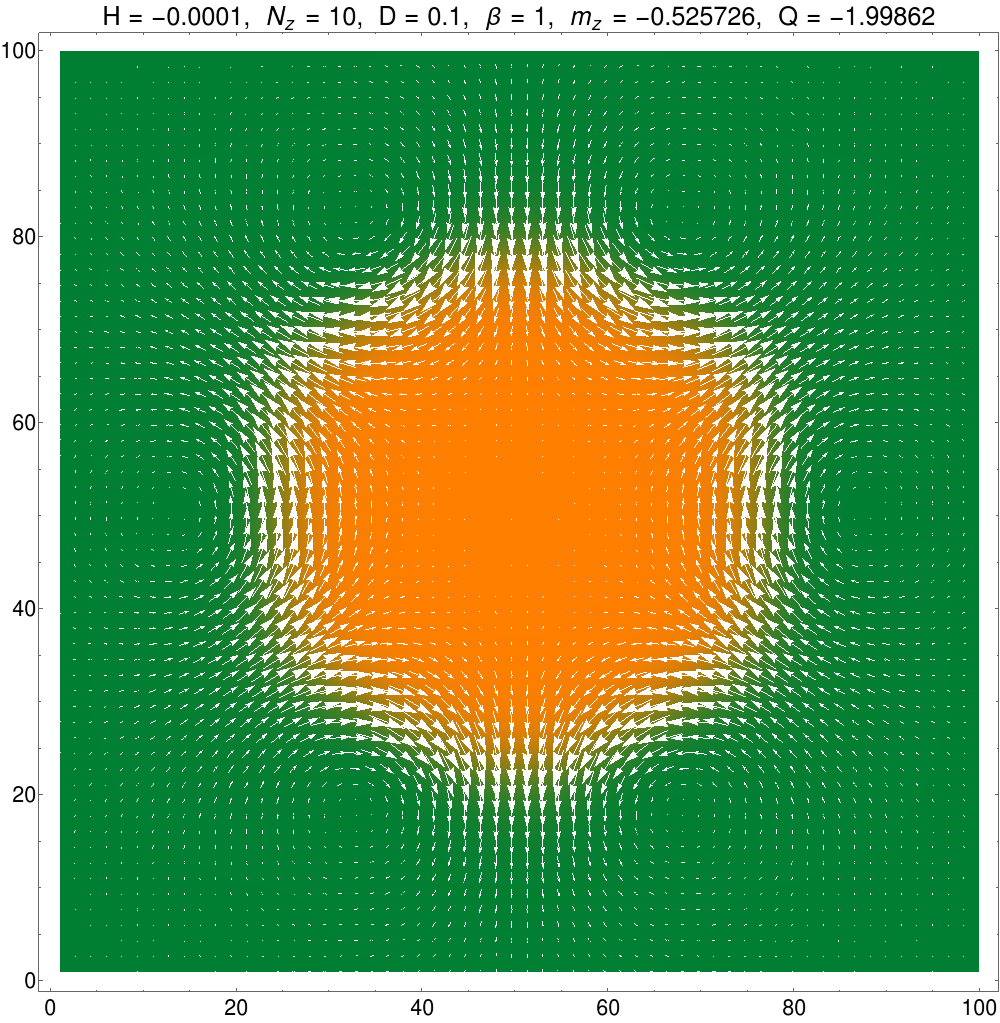}}
\centerline{\includegraphics[width=75mm]{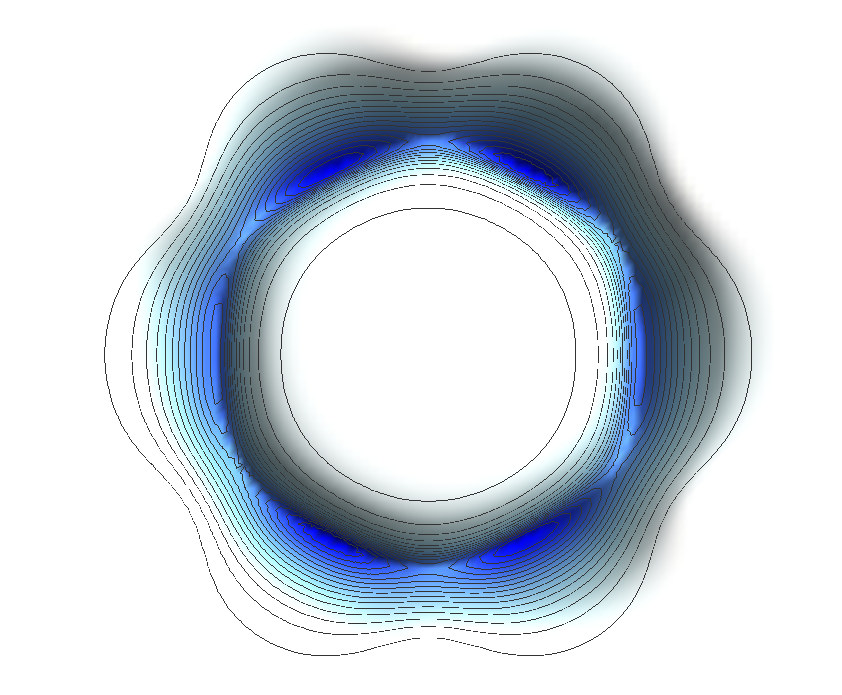}}
\vspace{0.5cm}
\caption{Antiskyrmion hexagon with $Q = -2$ in a $100\times 100\times 10$ film. (Upper panel) $m_z$; (Lower panel) density of the topological charge $q_S$. Red/blue codes positive/negative $q_S$. }
\label{hexagon}
\end{figure}

The heart-shaped cluster with $Q = 1$ shown in Fig.\ \ref{Combined-1} can be tentatively interpreted as two skyrmions coupled to an antiskyrmion. This cluster is unstable, however, since skyrmions and antiskyrmions tend to annihilate. As the field continues to increase, this eventually happens. Further increase of the field makes the remaining skyrmion to collapse, see movie in the supplemental material that illustrates the entire evolution of this cluster from zero to high field.

Further complex skyrmion objects evolving from magnetic bubbles with a uniform rotation of the in-plane magnetization in the wall as the initial state are shown in Figs.\ \ref{rosette} and \ref{hexagon}. As the result of the evolution, rotation localizes in relatively narrow regions of the wall -- Bloch lines, and the shape becomes a polygon or rosette. For any bubble, a simple connection exists between $Q$ and the total topological charge of the Bloch lines $N_B$: $Q = 1 + N_B/2$. By definition, $N_B=1$ if the in-plane magnetization in the wall rotates clockwise while moving clockwise along the wall, otherwise $N_B=-1$.  Notice that $N_B$ of the domain that gives rise to the skyrmion cluster is always even. In the case of well-localized BLs, the topological charge density $q_S$ has a peak at the BL making an integral contribution of $\pm 1/2$ into $Q$.

\begin{figure}
\centerline{\includegraphics[width=80mm]{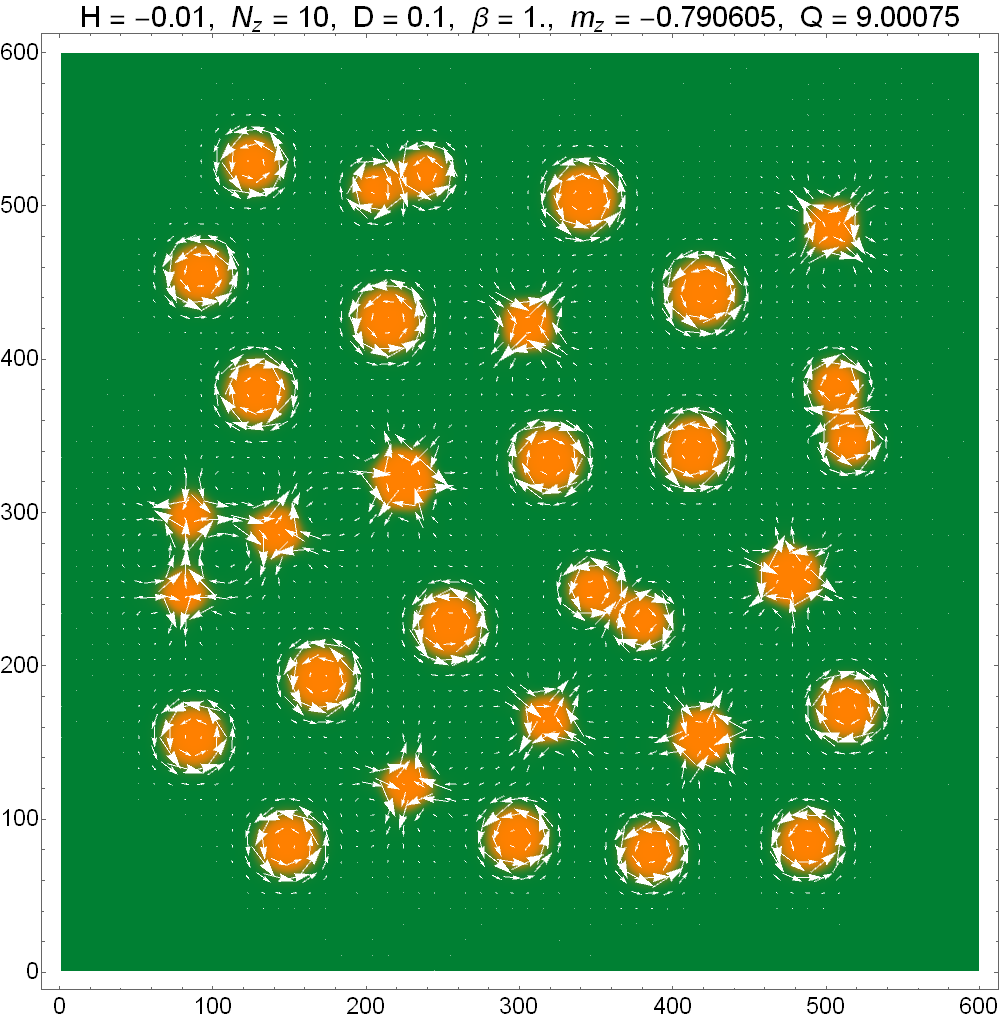}}
\centerline{\includegraphics[width=82mm]{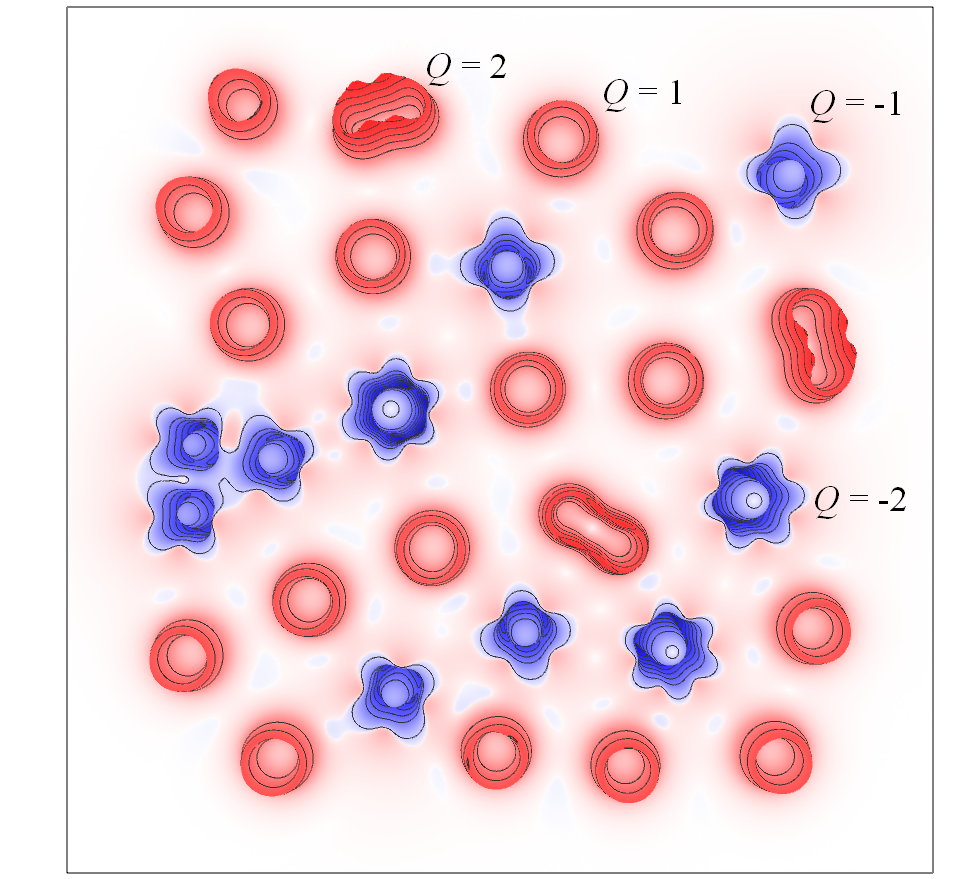}}
\vspace{0.5cm}
\caption{Skyrmion objects obtained by relaxation from the initial state with random direction of spins in the field $H=-0.01$ in a $600\times 600\times 10$ magnetic film (Upper panel) Perpendicular magnetization $m_z$ (orange/green) and  in-plane magnetization (white arrows); (Lower panel) The topological charge density.}
\label{Skyrmions_RIC}
\end{figure}

Skyrmion objects studied above play the role of building blocks in multiskyrmion patterns. One of them is shown in Fig.\ \ref{Skyrmions_RIC} obtained by relaxation in the perpendicular magnetic field $H=-0.01$ in a $600\times 600\times 10$ magnetic film starting from the initial state with random direction of spins. Similar states can be obtained by relaxation in zero field that yields a laminar domain structure with lots of Bloch walls (see lower panel of Fig.\ \ref{laminar}) and than increasing $H$ in the negative direction. For $H$ strong enough but not too strong, non-topological bubbles with $Q=0$ collapse and only topologically-protected objects remain. In the figure one can see 16 skyrmions ($Q=1$), 3 biskyrmions ($Q=2$), 9 antiskyrmions ($Q=-1$) and 3 hexagons ($Q=-2$). Three close antiskyrmions on the left are attracting via their apices and about to merge into an octagon ($Q=-3$). All antiskyrmions with different negative $Q$ have a regular polygon shape. Skyrmions show a strong tendency of grouping of Bloch lines in a small region of their wall, causing a large positive peak of the topological charge. In addition to biskyrmions, skyrmions with a
$180^\circ$
Bloch wall and $Q=2$ also arise naturally. Stronger magnetic field suppresses these BLs, transforming this object into a regular skyrmion. Rosettes with $Q=3$ and other objects with $Q>2$ were not detected in this numerical experiment.

Movies illustrating the dependence of the magnetic structure on the applied magnetic field can be found in Supplemental materials.

\section{Discussion}
We have shown that the variety of skyrmion objects observed in experiments on thin films has topological origin. It can be naturally explained by various combinations of Bloch lines in the domain walls of the laminar domain structure from which skyrmions emerge on the application of the magnetic field. Bloch lines are formed in a random statistical manner any time the magnetic domains are formed on either changing temperature or the magnetic field. In that sense the presence of Bloch lines in the domain walls is inevitable. The control over skyrmion structures evolving from laminar domains thus requires control of Bloch lines inside domain walls. Two additional observations are in order.

Firstly, the authors of Ref.\ \cite{Zhang2016} have recently found that the formation of multiskyrmion clusters is suppressed in samples that are annealed at elevated temperatures prior to magnetic measurements \cite{private}. This would be in line with the fact that pinning by defects of the crystal lattice stabilizes Bloch lines against moving inside domain walls towards the sample boundary. Suppression of the pinning by annealing allows Bloch lines to escape, thus diminishing the probability of the formation of a multiskyrmion cluster.

Secondly, the topological arguments presented in this paper would equally apply to magnetic bubbles that have been intensively studied in 1970s \cite{ODell}. The presence of Bloch lines inside bubble walls was known at that time. However, the multiskyrmion structures due to Bloch lines only emerge (on increasing the field) from stripe domains whose size is not very large compared to the domain wall width, that is, in very thin films with perpendicular magnetic anisotropy. Contemporary experimental studies of such systems have become possible due to the emergence of nanofabrication and nanoscale measuring techniques.

Thirdly, many compact topological objects that appear in experiments and called skyrmions often look like the objects shown in the upper panel of Fig.\ 9.  However, as is illustrated by the lower panel of Fig.\ 9, they are not necessarily simple skyrmions with $Q = \pm 1$. A more detailed study of the rotation of the magnetization inside these objects may be required to identify their true nature.

\acknowledgments
This work has been supported by the grant No. OSR-2016-CRG5-2977 from King Abdullah University of Science and Technology.


\begin{thebibliography}{0}
\bibitem{SkyrmePRC58} T. H. R. Skyrme, A non-linear theory of strong interactions, Proceedings of the Royal Society A \textbf{247}, 260-278 (1958).

\bibitem{Polyakov-book}
A. M. Polyakov, \textit{Gauge Fields and Strings}, Harwood Academic Publishers 1987.

\bibitem{BelPolJETP75} A. A. Belavin and A. M. Polyakov, Metastable states of two-dimensional isotropic ferromagnets, Pis'ma Zh. Eksp. Teor. Fiz \textbf{22}, 503-506 (1975) {[}JETP Lett. \textbf{22},
245-248 (1975); A. M. Polyakov, \textit{Gauge Fields and Strings}, Harwood Academic Publishers 1987.

\bibitem{Lectures}
E. M. Chudnovsky and J. Tejada, {\it Lectures on Magnetism}, Rinton Press (Princeton - NJ, 2006).

\bibitem{Brown-book}
Book: {\it The Multifaceted Skyrmion}, edited by G. E. Brown and M. Rho (World Scientific, 2010).

\bibitem{AlkStoNat01} U. Al'Khawaja, and H. T. C. Stoof, Skyrmions in a ferromagnetic Bose-Einstein condensate, Nature \textbf{411}, 918-920 (2001).

\bibitem{SonKarKivPRB93} S. L. Sondhi, A. Karlhede, S. A. Kivelson,
and E. H. Rezayi, Skyrmions and the crossover from the integer to fractional quantum Hall effect at small Zeeman energies, Physical Review B \textbf{47}, 16419-16426 (1993).

\bibitem{YeKimPRL99} Jinwu Ye, Y. B. Kim, A. J. Millis, B. I. Shraiman,
P. Majumdar, and Z. Tesanovic, Berry phase theory of the anomalous Hall effect: Application to colossal magnetoresistance manganites, Physical Review Letters \textbf{83}, 3737-3740 (1999).

\bibitem{StonePRB93} M. Stone, Magnus force on skyrmions in ferromagnets and quantum Hall systems, Physical Review B \textbf{53}, 16573-16578 (1996).

\bibitem{WriMerRMR89} D. C. Wright, and N. D. Mermin, Crystalline liquids: the blue phases, Review of Modern Physics \textbf{61}, 385-433 (1989).

\bibitem{CCG}
L. Cai, E. M. Chudnovsky, and D. A. Garanin, Collapse of skyrmions in two-dimensional ferromagnets and antiferromagnets, Physical Review B {\bf 86}, 024429-(4) (2012).

\bibitem{Koshibae2016}
W. Koshibae and N. Nagaosa, Theory of antiskyrmions in magnets, Nature Communications {\bf 7}, 10542-(8) (2016).

\bibitem{Nagaosa2013}
N. Nagaosa and Y. Tokura, Topological properties and dynamics of magnetic skyrmions, Nature Nanotechnology {\bf 8}, 899-911 (2013).

\bibitem{Klaui2016}
G. Finocchio, F. B\"{u}ttner, R. Tomasello, M. Carpentieri, and M. Klaui, Magnetic skyrmions: from fundamental
to applications, Journal of Physics D: Applied Physics {\bf 49}, 423001-(17) (2016).

\bibitem{Yu-2014}
X. Z. Yu, Y. Tokunaga, Y. Kaneko, W. Z. Zhang, K. Kimoto, Y. Matsui, Y. Taguchi, and  Y. Tokura, Biskyrmion states and their current-driven motion in a layered manganite, Nature Communications {\bf 5}, 3198-(7) (2014).

\bibitem{Zhang2016}
W. Wang, Y. Zhang, G. Xu, L. Peng, B. Ding, Y. Wang, Z. Hou, X. Zhang, X. Li, E. Liu, S. Wang, J. Cai, F. Wang, J. Li, F. Hu, G. Wu, B. Shen, and X.-X. Zhang, A centrosymmetric hexagonal magnet with superstable biskyrmion magnetic nanodomains in a wide temperature range of $100-340$K, Advanced Materials {\bf 28}, 6887-6893 (2016).

\bibitem{PNAC2016}
X. Zhao, C. Jin, C. Wang, H. Du, J. Zang, M. Tian, R. Che, and Y. Zhang, Direct imaging of magnetic field-driven transitions of skyrmion cluster states in FeGe nanodisks, Proceedings of the U.S. National Academy of Sciences {\bf 113}, 4918-4923 (2016).

\bibitem{Muller2017}
J. M\"{u}ller, J. Rajeswari, P. Huang, H. M. Ronnow, F. Carbone, and A. Rosch, Magnetic skyrmions and skyrmion clusters in the helical phase of Cu$_2$OSeO$_3$, cond-mat arXiv:1703.06989 (2017).

\bibitem{Loudon2017}
J. C. Loudon, A. O. Leonov, A. N. Bogdanov, M. Ciomaga Hatnean, and G. Balakrishnan, Direct observation of attractive skyrmions and skyrmion clusters in a cubic helimagnet Cu$_2$OSeO$_3$, cond-mat arXiv:1704.06876 (2017).

\bibitem{Leonov-NJP2016}
A. O. Leonov, T. L. Monchesky, N. Romming, A. Kubetzka, A. N. Bogdanov, and R. Wiesendanger, The properties of isolated chiral skyrmions in thin magnetic films, New Journal of Physics {\bf 18}, 065003-(16) (2016).

\bibitem{Leonov-JPCM2016}
A. O. Leonov, T. L. Monchesky, J. C. Loudon, A. N. Bogdanov, Three-dimensional chiral skyrmions with attractive interparticle interactions, Journal of Physics: Condensed Matter {bf 28}, 35LT01-(5) (2016).

\bibitem{Leonov-NatCom2015}
A. O. Leonov and M. Mostovoy, Multiply periodic states and isolated skyrmions in an anisotropic frustrated magnet, Nature Communications {\bf 6}, 8275-(8) (2015).

\bibitem{NNN-2017}
X. Zhang, J. Xia, Y. Zhou, X. Liu, and M. Ezawa, Skyrmions and antiskyrmions in a frustrated $J_1-J_2-J_3$ ferromagnetic film: Current-induced helicity locking-unlocking transition, cond-mat arXiv:1703.07501 (2017).

\bibitem{garchupro13prb}
D. A. Garanin, E. M. Chudnovsky, and T. Proctor, Random field $xy$ model in three dimensions, Physical Review B {\bf 88},  224418-(21) (2013).

\bibitem{private}
Z. Hou, private communication.

\bibitem{ODell}
T. H. O'Dell, {\it Ferromagnetodynamics: The Dynamics of Magnetic Bubbles, Domains, and Domain Walls}, Wiley 1981.



\end{thebibliography}
\end{document}